\documentclass[a4paper,11pt]{article}
\usepackage{pos}

\usepackage{bm}
\usepackage{cancel}
\usepackage{braket}
\usepackage{url}

\title{Investigation of the baryon time-like electromagnetic form factors in the electron-positron annihilation reactions}

\author*{Ju-Jun Xie}
\author{Cheng Chen}

\affiliation{Heavy Ion Science and Technology Key Laboratory, Institute of Modern Physics, Chinese Academy of Sciences, Lanzhou 730000, China}
\affiliation{Southern Center for Nuclear-Science Theory (SCNT), Institute of Modern Physics, Chinese Academy of Sciences, Huizhou 516000, China}
\affiliation{School of Nuclear Sciences and Technology, University of Chinese Academy of Sciences, Beijing 101408, China}

\emailAdd{xiejujun@impcas.ac.cn}
\emailAdd{chencheng@impcas.ac.cn}

\abstract{
Based on the experimental measurements of the electron-positron annihilation reactions into a baryon ($B$) and anti-baryon ($\bar{B}$) pair, the electromagnetic form factors of hyperons in the time-like region can be investigated within the vector meson dominance model. The theoretical model parameters are determined by fitting them to the total cross sections of the process $e^+e^-\to B \bar{B}$,  and it is found that the current experimental data on the baryon electromagnetic form factors in the time-like region can be well reproduced. In addition to the total cross sections, the electromagnetic form factors $G_E$ and $G_M$, and the charge radii of those baryons are also estimated, which are in agreement with the experimental data. On the other hand, we have also investigated the nonmonotonic behavior of the time-like baryon electromagnetic form factors, and it is found that the previously proposed periodic behaviour of the nucleon time-like electromagnetic form factor is not confirmed. However, we do observe the nonmonotonic structures in the line shape of the baryon effective form factors or the ratio $|G_E/G_M|$ for the charmed baryon $\Lambda^+_c$. These features can be naturally explained by incorporating contributions from excited vector states. More precise measurements of the $e^+e^-\to B \bar{B}$ reaction offer a valuable opportunity to probe the properties of excited vector states, which are at present poorly known. Additionally, comprehensive theoretical and experimental studies of baryon timelike electromagnetic form factors can provide critical insights into the underlying mechanisms of electron-positron annihilation processes.
}

\FullConference{The XVIth Quark Confinement and the Hadron Spectrum Conference (QCHSC24)\\
 19-24 August, 2024\\
 Cairns Convention Centre, Cairns, Queensland, Australia\\}

\begin{document}
\maketitle

\section{Introduction}

In classical quark models, baryons are composite particles consisting of three valence quarks ($qqq$) and a neutral sea of strong interaction. Because of the  non-perturbative nature of strong interaction in the energy regime of the low-lying baryons, an exact theoretical description of baryon internal structure has not been achieved within the framework of quantum chromodynamics theory. The baryon electromagnetic structure can be described by the electromagnetic form factors (EMFFs), which depend on the squared four-momentum ($q^2$) of coupled virtual photon. Investigating the EMFFs of baryons is essential for gaining deeper insight into their fundamental structure~\cite{Denig:2012by,Pacetti:2014jai}. In the space-like region ($q^2 < 0$), the EMFFs are real and they can be measured, in general, from $e^- B \to e^- B$ elastic scattering. The measurements of space-like region EMFFs of proton can be done in elastic as well as inelastic $ep$ scattering. However, for other unstable baryons, since their lifetimes are so short and thus they can not used as target, measuring their EMFFs at the space-like region is very challenging. Instead, the electron-position annihilation process, $e^+ e^- \to B \bar{B}$ allows us to study the baryon EMFFs at the time-like region. At a fixed-energy of $e^+e^-$ collision, the baryon EMFFs in the time-like region were extracted from the data on the differential cross section of the process $e^+e^- \to B \bar{B}$. Consequently, a new era has begun with the introduction of electron-positron annihilation reactions~\cite{Ping:2013jka,BESIII:2017lkp}, where a baryon-antibaryon pair is formed by a virtual photon~\footnote{We have assumed that the one-photon exchange is dominant.}. Indeed, on the experimental side, significant progress has been made in measuring baryon EMFFs in time-like region~\cite{BaBar:2007fsu,BESIII:2017hyw,BESIII:2019nep,BESIII:2019cuv,BESIII:2020uqk,BESIII:2021aer,BESIII:2021rkn,Belle:2022dvb,BESIII:2023ldb}. 

On the theoretical side, the vector meson dominance (VMD) model is a successful approach for studying the baryon EMFFs, in both space-like and time-like regions~\cite{Iachello:1972nu,Iachello:2004aq,Bijker:2004yu}. In the studying baryon electromagnetic form factors with the VMD model, there is a phenomenological intrinsic form factor $g(q^2)$, which is a characteristic of valence quark structure. From these investigations of the nucleon and hyperon EMFFs~\cite{Iachello:1972nu,Iachello:2004aq,Bijker:2004yu,Yang:2019mzq}, it is found that a better choice of $g(q^2)$ is the dipole form, $g(q^2) = 1/(1-\gamma_B q^2)^2$, with $\gamma_B$ a free model parameter. And, the $e^+ e^- \to B \bar{B}$ reaction is a very good flat to study the excited vector mesons that couples to the $B\bar{B}$ channel~\cite{Li:2020lsb,Dai:2021yqr,Yan:2023nlb,Chen:2023oqs,Chen:2024luh}.

In this study, we investigate the time-like electromagnetic form factors for nucleon, $\Lambda$, $\Sigma$, $\Xi$, and $\Lambda^+_c$, using the extended vector meson dominance model.

\section{Theoretical framework}

Based on parity conservation and Lorentz invariance, the electromagnetic current of the baryons with a spin of $1/2$ can be characterized by two independent scalar functions $F_1(q^2)$ and $F_2(q^2)$ depending on $q^2$, which are the Dirac and Pauli form factors, respectively. Then the corresponding electric and magnetic form factors $G_E(q^2)$ and $G_M(q^2)$ are written as~\cite{Pacetti:2014jai},
\begin{eqnarray}
	G_E(q^2) = F_1(q^2) + \tau F_2(q^2), ~~~ ~~~	G_M(q^2) = F_1(q^2) + F_2(q^2),
\end{eqnarray}
where $M$ is the baryon mass and $\tau = q^2/(4M^2)$. With $G_E(q^2)$ and $G_M(q^2)$, the effective form factor $|G_{\rm eff}(q^2)|$ is defined as
\begin{eqnarray}
	| G_{\rm eff} (q^2) | = \sqrt{\frac{2\tau|G_M(q^2)|^2 + | G_E(q^2)|^2}{1+2\tau}}.
\end{eqnarray}
 The module squared of effective form factor $|G_{\rm eff}|^2$ is a linear combination of $|G_E|^2$ and $|G_M|^2$, and proportional to the total cross section of $e^+e^- \to B \bar{B}$ reaction. It also indicates how much the experimental $e^+ e^- \to B \bar{B}$ cross section differs from a point-like baryon $B$.

\begin{figure}[htbp]
	\centering
	\includegraphics[scale=0.8]{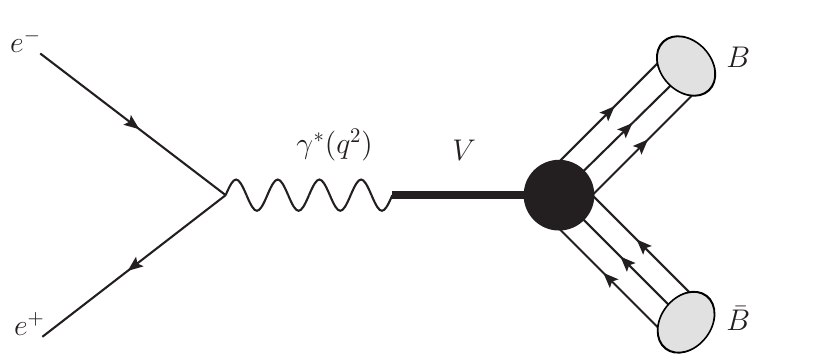}
	\caption{The Feynman diagram of $e^+e^-\rightarrow B \bar{B}$ in the VMD model.} 
	\label{fig:F}
\end{figure}

The effective form factor $|G_{\mathrm{eff}}|$ can be extracted from the Born cross section of $e^+e^-\to B \bar{B}$ reaction, which represents the dependence of effective coupling strength of the photon-baryon interaction vertex $\gamma^* B\bar{B}$. In the VMD model, the virtual photon coupling to baryons is through the intermediate vector mesons, as shown in Fig.~\ref{fig:F}, where $V$ stands the vector mesons that have significant couplings to the final states $B\bar{B}$. Thus the Dirac and Pauli form factors are parameterized as follows:
\begin{eqnarray}
	F_1 (q^2) = g(q^2)\left( f_1 +  \sum_{i=1}^n \beta_i B_{R_i}  \right ),  ~~~~~~
	F_2 (q^2) = g(q^2)\left( f_2 B_{R_1}  + \sum_{i=2}^n \alpha_i B_{R_i} \right), 
\end{eqnarray}
with 
\begin{eqnarray}
	B_{R_i} = \frac{M_{R_i}^2}{M_{R_i}^2 - q^2 - i M_{R_i} \Gamma_{R_i}}, ~~~	f_1 = C_B - \beta_1 - \beta_2 - \beta_3 - \beta_4 , ~~~ f_2 = \mu_{B} - C_B - \sum_{i=2}^n \alpha_i ,
\end{eqnarray}
where $M_{R_i}$ and $\Gamma_{R_i}$ are the mass and width of the excited vector state $R_i$. And, $C_B = 1$ or $0$ for charged 1 or neutral baryon, respectively. $\mu_B$ is the value of baryon magnetic moment in natural unit, i.e., $\hat{\mu}_B = e/(2M_B)$. The $\beta_i$ and $\alpha_i$ are theoretical model parameters, which are fitted to the experimental data on the $e^+ e^- \to B\bar{B}$ reaction.

\section{Numerical results and discussions}

\subsection{The $e^+ e^- \to N \bar{N}$ reaction}

\begin{figure}[htbp]
	\centering
	\includegraphics[scale=0.4]{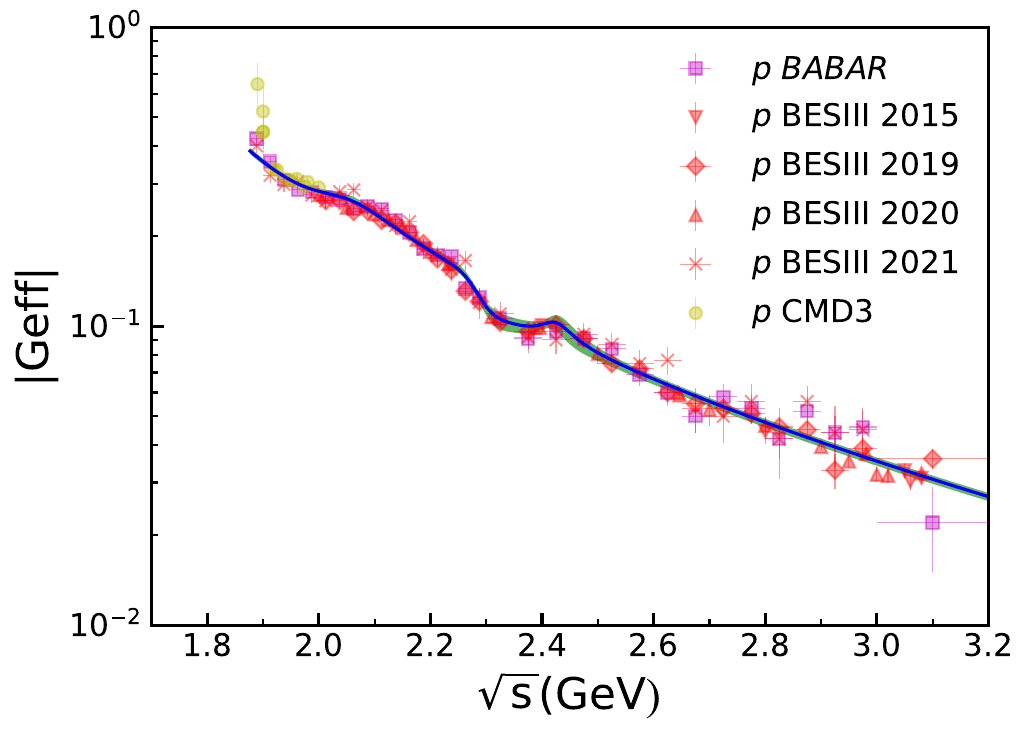}
	\includegraphics[scale=0.4]{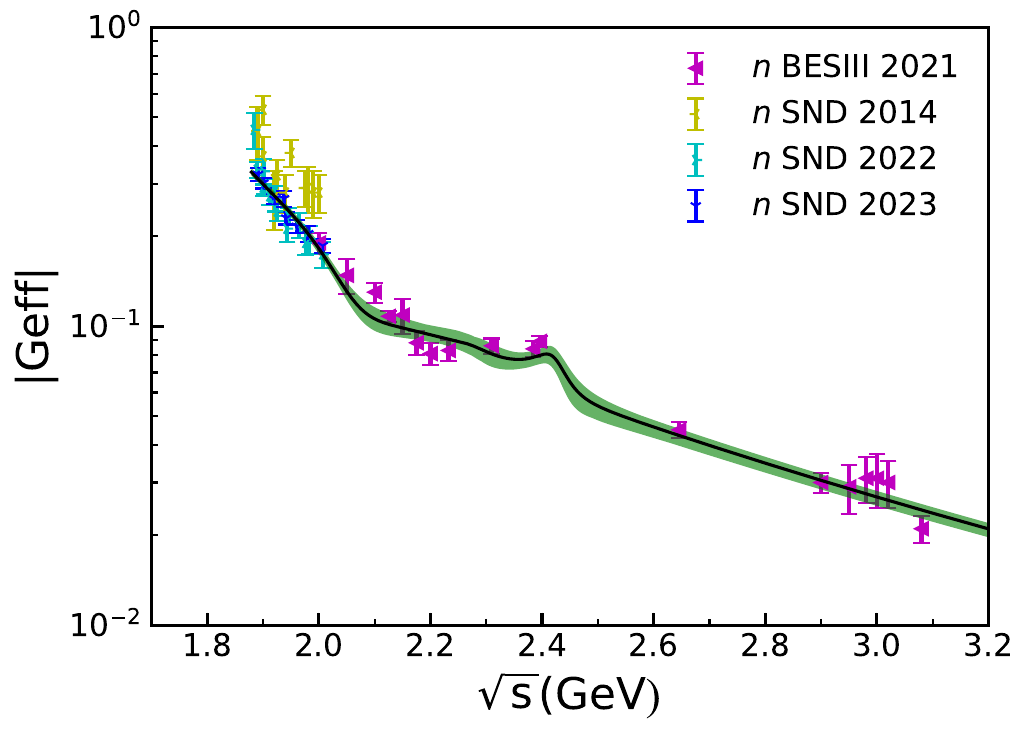}
\caption{The fitted results of  the effective form factors for proton (left) and neutron (right), comparing with experimental data taken from: $BABAR$~\cite{BaBar:2013ves}, BESIII 2015~\cite{BESIII:2015axk}, BESIII 2019~\cite{BESIII:2019tgo}, BESIII 2020~\cite{BESIII:2019hdp}, BESIII 2021~\cite{BESIII:2021rqk} for proton, CMD3~\cite{CMD-3:2015fvi}, BESIII 2021~\cite{BESIII:2021tbq} for neutron, SND 2014~\cite{Achasov:2014ncd}, SND 2022~\cite{SND:2022wdb}, and SND 2023~\cite{SND:2023fos}.}   \label{fig:EMFFs}
\end{figure}

We show firstly the theoretical numerical results for the effective form factors of proton and neutron in Fig.~\ref{fig:EMFFs}, comparing with the experimental data. The theoretical error bands are obtained with the uncertainties of fitted parameters. It is found that both proton and neutron effective form factors can be well reproduced including the contributions from the excited $\rho(2D)$, $\omega(3D)$, and $\omega(5S)$ states~\cite{Yan:2023nlb}. This indicates that the nonmonotonic line shapes of the nucleon effective form factors can be explained within the VMD model, where the contributions of excited vector mesons are taken into account. The numerical results show that the so-called periodic behaviour of the nucleon effective form factors is not confirmed. However, there are indeed nonmonotonic structures in the line shape of nucleon effective form factors, which can be naturally reproduced by considering the contributions from the low-lying excited vector states.

\subsection{The $e^+ e^- \to \Lambda {\bar\Lambda}$ reaction}

\begin{figure}[htbp]
	\centering
	\includegraphics[scale=0.5]{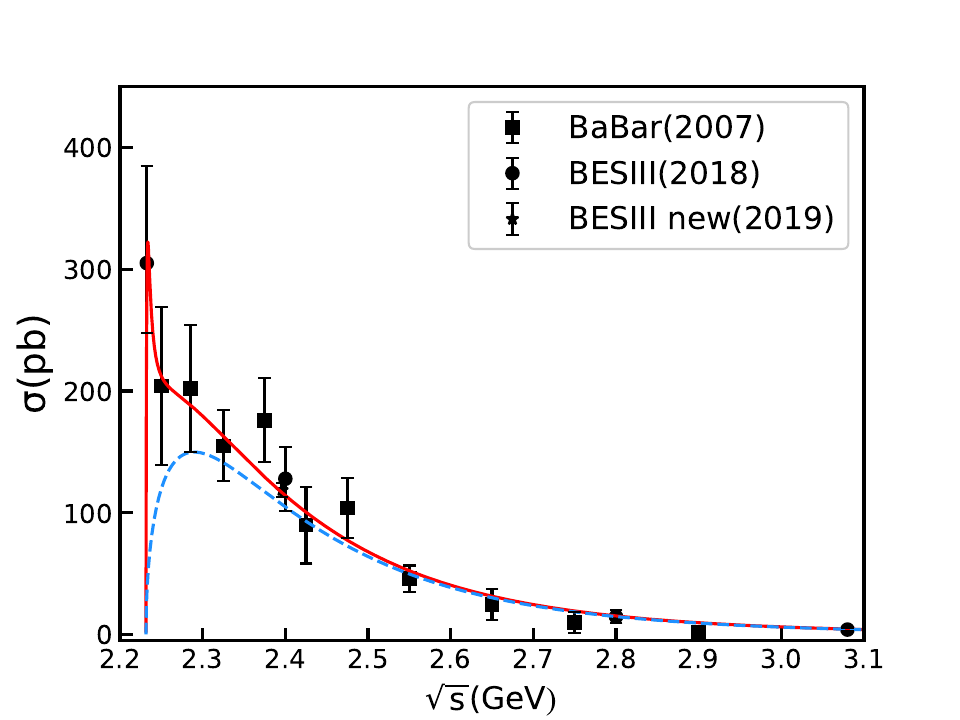}
	\caption{The total cross section of $e^+ e^- \to \Lambda\bar{\Lambda}$ reaction compared with the experimental data measured by BABAR collaboration~\cite{BaBar:2007fsu} and BESIII collaboration~\cite{BESIII:2017hyw,BESIII:2019nep}.} \label{fig:tcsLambda}
\end{figure}

The new experimental measurements of the BESIII collaboration~\cite{BESIII:2017hyw,BESIII:2019nep} indicates there is a large threshold
enhancement for the $e^{+}e^{-} \to \Lambda \bar{\Lambda}$ reaction. The observed value close to the reaction threshold is larger than the previous theoretical predictions. In Fig.~\ref{fig:tcsLambda}, the theoretical fitted results of the $e^+e^- \to \Lambda\bar{\Lambda}$ total cross sections in the energy range from the reaction threshold to $\sqrt{s} = 3.1$ GeV are shown and compared to experimental data taken from Refs.~\cite{BESIII:2017hyw,BESIII:2019nep}. One can see that the near threshold enhancement structure is fairly well reproduced thanks to a significant contribution from a very narrow vector meson $X(2232)$ with a mass of about 2231.8 MeV, and it has significant couplings to the $\Lambda\bar{\Lambda}$ channel. The narrow peak of this state is clearly seen. This solves the problem that all previous calculations seriously underestimate the near-threshold total cross section of the $e^+e^- \to \Lambda\bar{\Lambda}$ reaction. However, the width of the state cannot be well determined through the VMD model by fitting the current experimental data~\cite{Li:2021lvs}.

\section{The $e^+e^- \to \Sigma\bar{\Sigma}$ reaction}

\begin{figure}[htbp]
	\centering
	\includegraphics[scale=0.35]{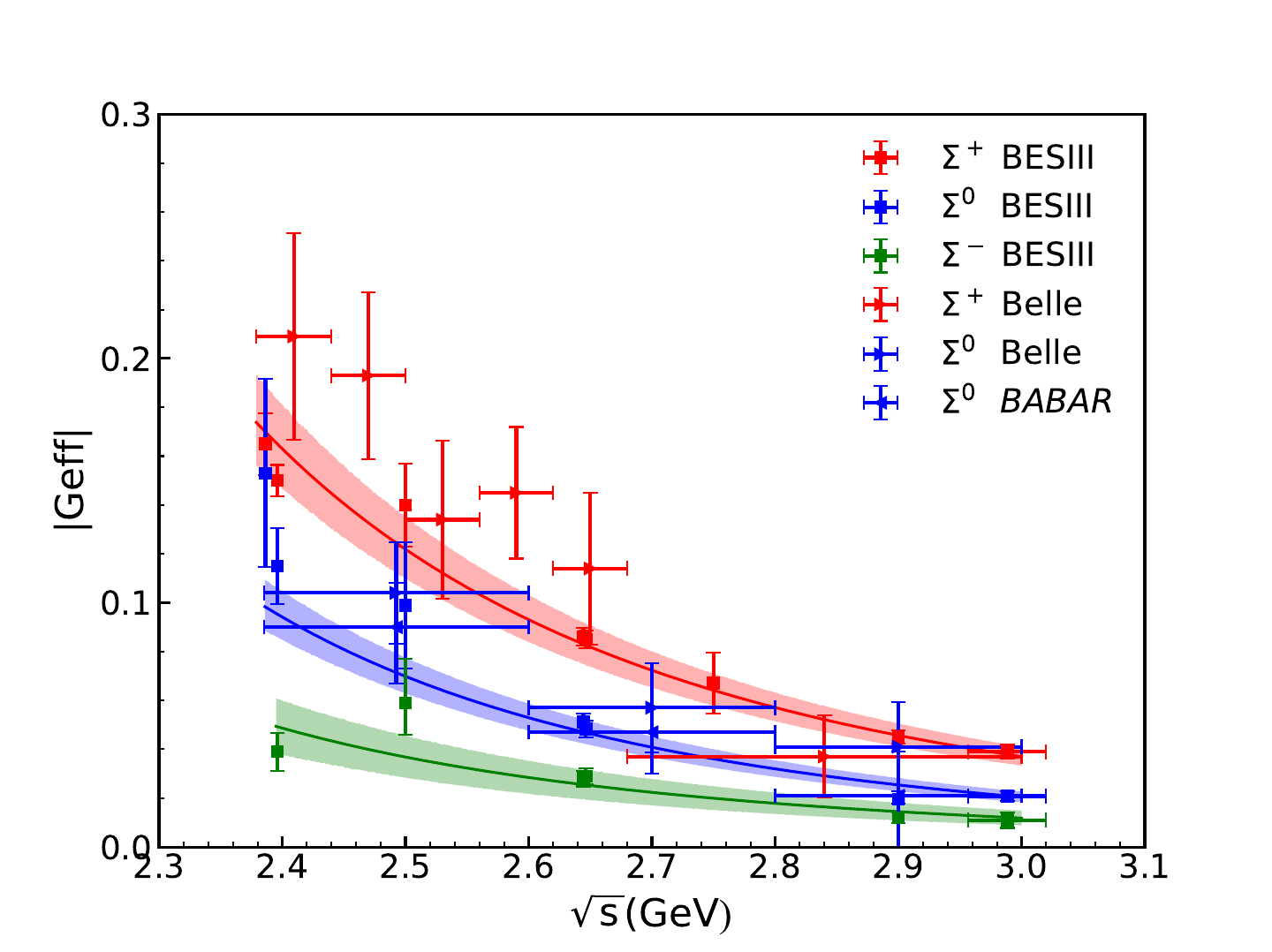}
	\caption{The obtained effective form factors of $\Sigma^+$, $\Sigma^0$, and $\Sigma^-$, compared with the experimental data.}	\label{fig:GeffSigma}
\end{figure}

In Fig.~\ref{fig:GeffSigma} we show the theoretical results of the effective form factors of the $\Sigma^+$, $\Sigma^0$, and $\Sigma^-$. The red, blue, and green curves stand for the results for $\Sigma^+$, $\Sigma^0$, and $\Sigma^-$, respectively. The band accounts for the corresponding
$68\%$ confidence-level interval deduced from the distributions of the fitted parameters~\cite{Yan:2023yff}. The experimental data from BESIII~\cite{BESIII:2020uqk,BESIII:2021rkn}, Belle~\cite{Belle:2022dvb}, and $BABAR$ Collaboration~\cite{BaBar:2007fsu} are also shown for comparison. One can see that, with the same model parameters, we can describe these data on the effective form factors of $\Sigma^+$, $\Sigma^0$ and $\Sigma^-$ quite well. The resulting ratios of total cross sections of these above three reactions are $9.7 \pm 1.3$ : $1$ : $3.3 \pm 0.7$~\cite{BESIII:2020uqk}, which disagree with various theoretical model predictions, can be also naturally explained here.

\section{The $e^+e^- \to \Xi \bar{\Xi}$ reaction}

\begin{figure}[htbp]
	\centering
	\includegraphics[scale=0.35]{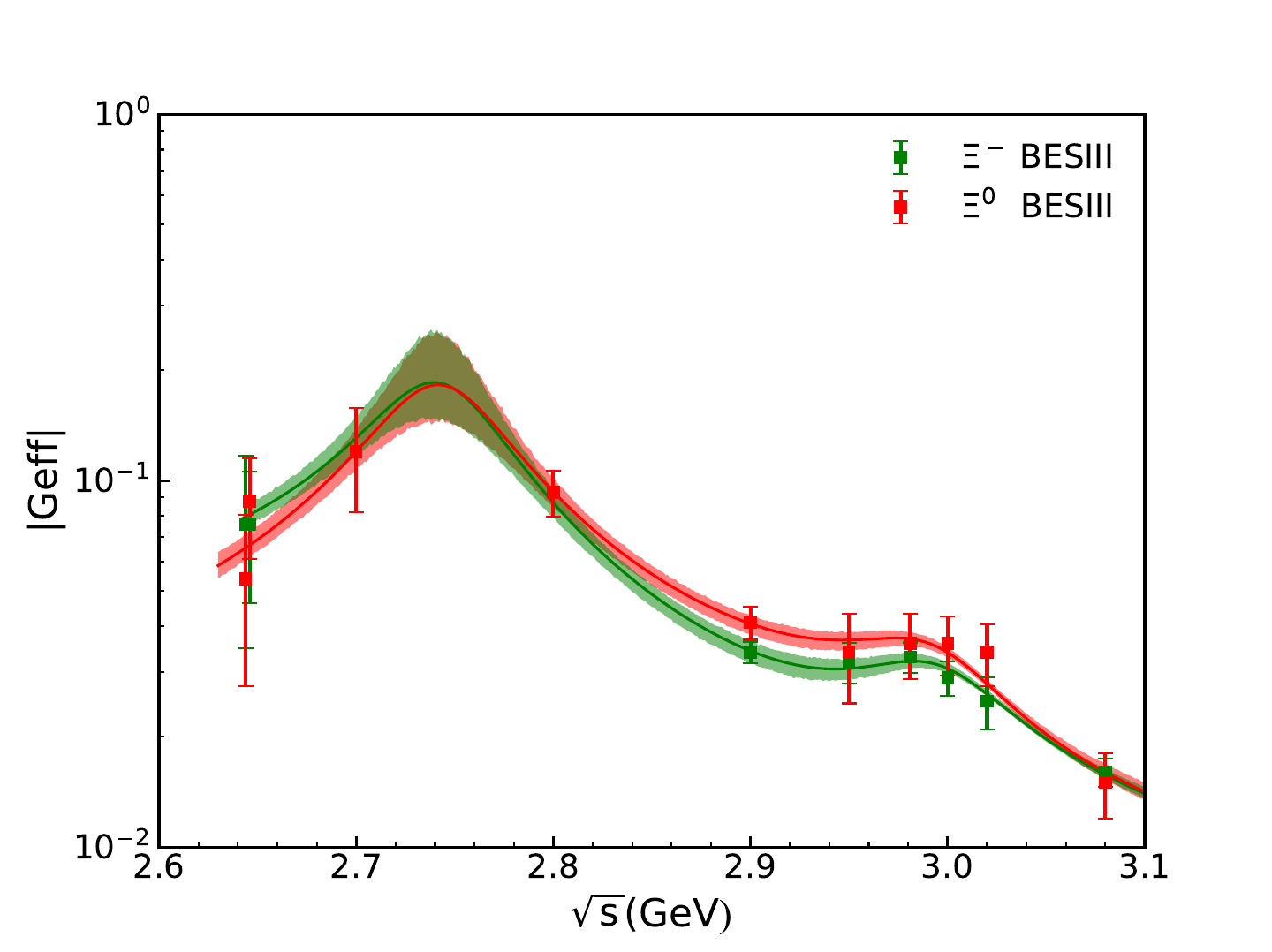}
	\caption{The obtained $\Xi^-$ and $\Xi^0$ effective form factors compared with the experimental data.} \label{fig:GeffXi}
\end{figure}

For the case of $\Xi^-$ and $\Xi^0$ effective form factors, the numerical results are shown in Fig.~\ref{fig:GeffXi}, where the red curve stands for the results of $\Xi^0$, while the green curve is the fitted results for $\Xi^-$. We also show the statistical error bands for the fitted results. Again, one can see that the experimental data on the effective form factors of $\Xi^-$ and $\Xi^0$ can be well reproduced. It is worth mentioning that the two resonances $V_1$ and $V_2$ are crucial to describe the experimental data, and without their contributions, we cannot get a good fit for the experimental data (see more details in Ref.~\cite{Yan:2023yff}). Their masses and widths are: $M_{V_1} = 2.742$ GeV, $\Gamma_{V_1} = 71$ MeV, $M_{V_2} = 2.993$ GeV, and $\Gamma_{V_2} =  88$ MeV. It is expected that new precise experimental data at BESIII~\cite{BESIII:2020nme} can be used to further study their properties. Especially, more data points around the $V_1$ peak are crucial to check the importance of the new vector state $V_1$.

\section{The $e^+e^- \to \Lambda^+_c \bar{\Lambda}^-_c$ reaction}

\begin{figure}[htbp]
	\centering
	\includegraphics[scale=0.3]{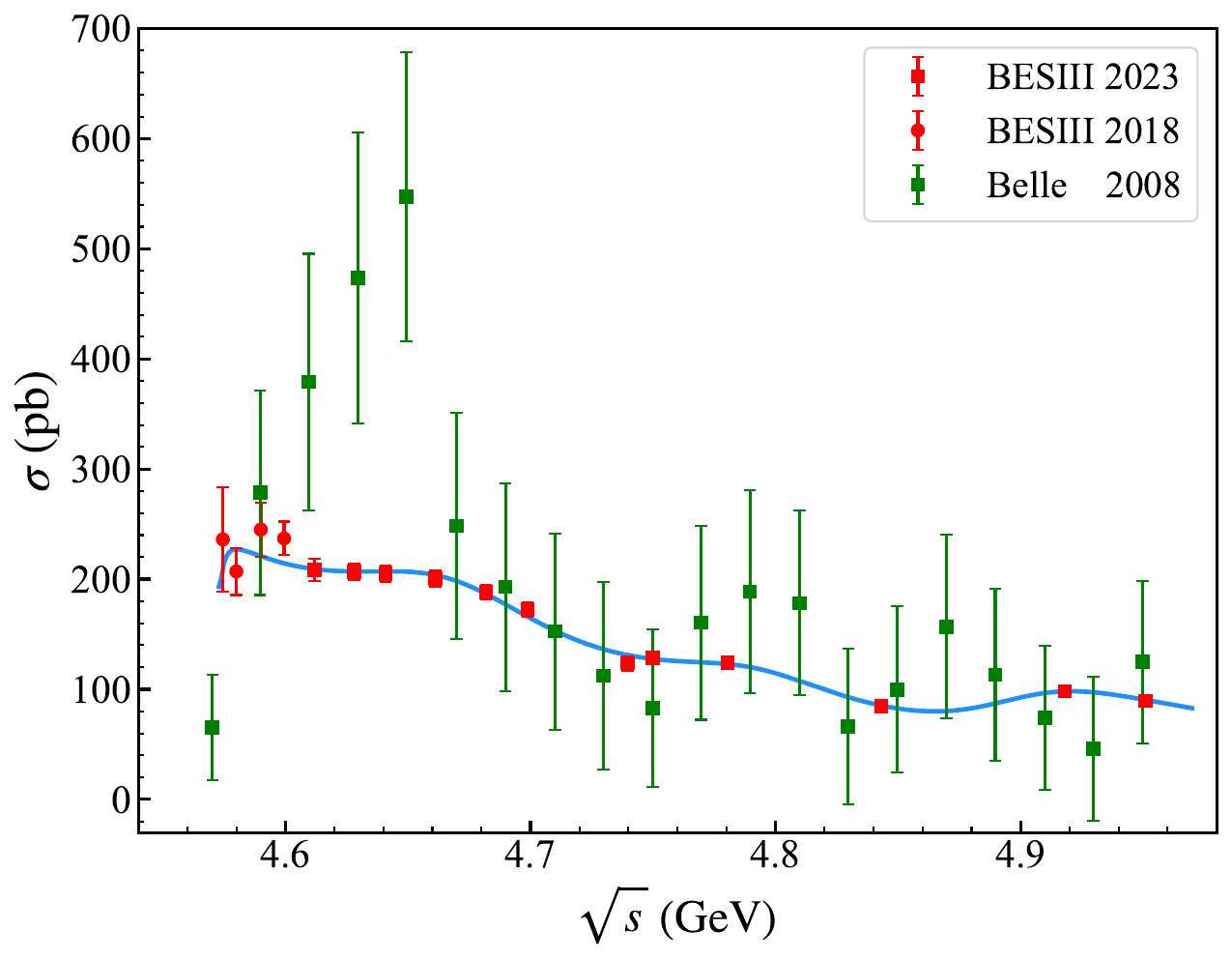}
    \includegraphics[scale=0.3]{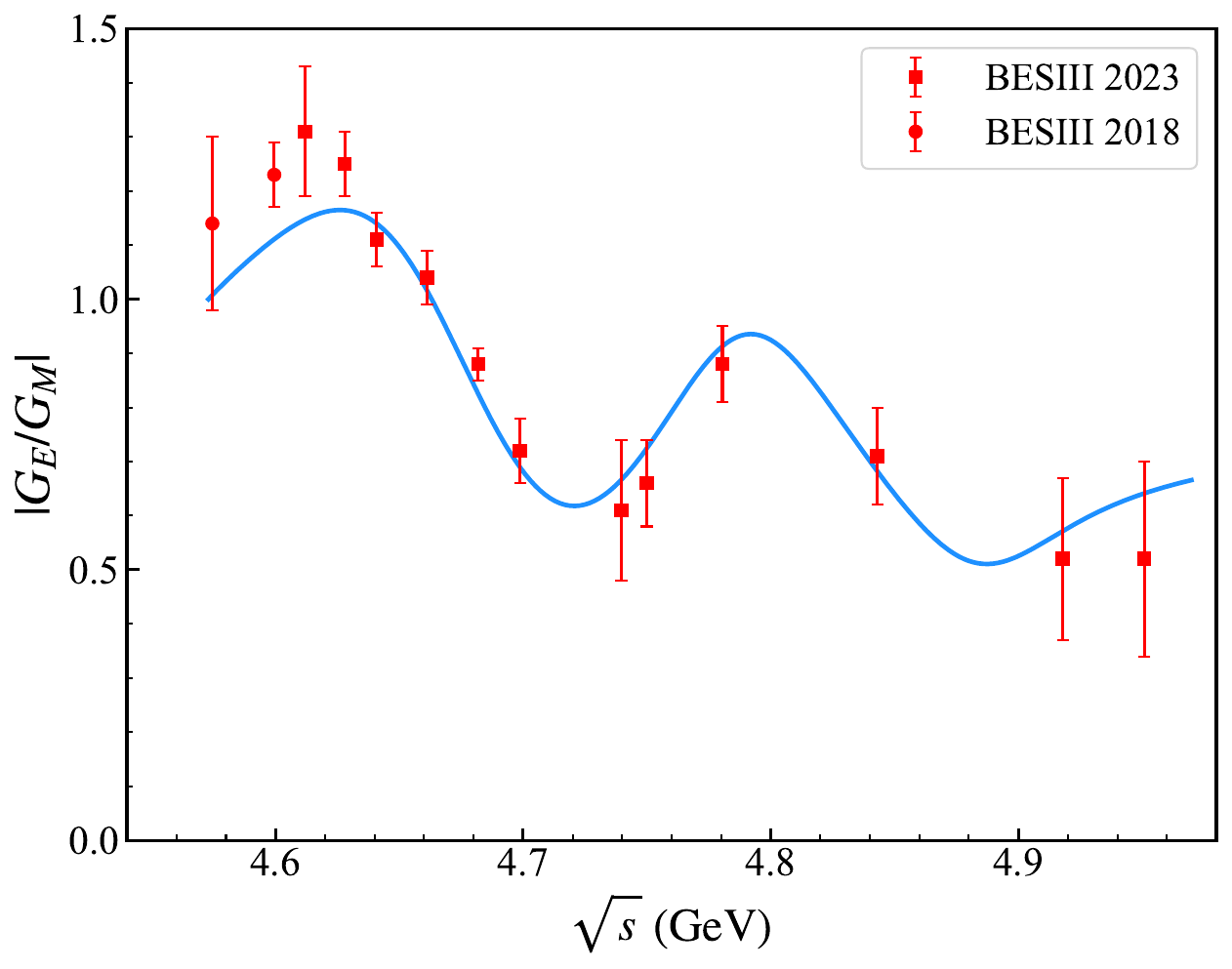}
	\caption{The obtained total cross sections (left) and ratio $|G_E/G_M|$ compared with the experiment data taken from BESIII 2018~\cite{BESIII:2017kqg}, BESIII 2023~\cite{BESIII:2023rwv}, and Belle 2008~\cite{Belle:2008xmh}.}
	\label{fig:TCSandRatio}
\end{figure}

In Fig.~\ref{fig:TCSandRatio}, the fitted results for the total cross sections of $e^+e^- \to \Lambda^+_c \bar{\Lambda}^-_c$ reaction and the ratio $|G_E/G_M|$ are shown and compared with the experimental data. Note that the Belle Collaboration data were not taken into account in the $\chi^2$ fit. One can see that, within the VMD model and by considering the contributions from $\psi(4500)$, $\psi(4660)$, $\psi(4790)$, and $\psi(4900)$, the total cross sections of $e^+e^- \to \Lambda^+_c \bar{\Lambda}^-_c$ reaction can be well described in the considered energy regions $4.57 \ \mathrm{GeV}< \sqrt{s} < 4.96\ \mathrm{GeV} $.  Although the structure of $\psi(4660)$ is not significant, the contribution from it is essential~\cite{Chen:2023oqs,Chen:2024luh}. Similarly, in the higher energy region, the cross sections from $4.74 \ \mathrm{GeV} $ to $4.79 \ \mathrm{GeV} $ shows a plateau with an approximate width of $50 \ \mathrm{MeV} $, attributed to the two states of $\psi(4660)$ and $\psi(4790)$. Moreover, the $\psi(4900)$ state is needed to explain the small bump structure around $4.92$ GeV.

 It is found that the obtained $|G_E/G_M|$ are in good agreement with experimental data. The non-monotonic structures (or the so-called oscillating behavior as in Ref.~\cite{BESIII:2023rwv}) shown in the line shape of the ratio $|G_E/G_M|$ can be naturally explained by including the contributions from the charmonium states. And the $\psi(4900)$ state is crucial to reproduce the experimental measurements on the $e^+e^- \to \Lambda_c^+ \bar{\Lambda}_c^-$ reaction, even there are a few data points in the energy regions from $4.85\ \mathrm{GeV}$ to $4.96\ \mathrm{GeV}$. It is expected that more experimental measurements around $4.9$ GeV can be used to further study the possible $\psi(4900)$ state~\cite{Jia:2023upb}.

\section{Summary and conclusions}
	
In this study, the baryon electromagnetic form factors in the time-like region were investigated within the extended vector meson dominance model. In addition to these contributions from ground vector mesons, we introduce also excited vector states. The model parameters are determined by fitting them to the experimental data on the $e^+e^- \to B \bar{B}$ reaction. Using the fitted theoretical parameters, the effective form factor can be calculated, and it is found that the theoretical results are also consistent with the experimental data. Moreover, we conclude that the non-monotonic structures observed in the line shape of the $e^+ e^- \to B\bar{B}$ total cross sections can be naturally explained within the vector meson dominance model. Additionally, the $e^+ e^- \to B\bar{B}$ reactions can be used to study the excited vector states, especially for those exotic states that couple strongly to the $B\bar{B}$ channels.

\section*{Acknowledgments}

We would like to thank Bing Yan, An-Xin Dai, and Zhong-Yi Li for collaboration on relevant issues presented here. This work is partly supported by the National Key R\&D Program of China under Grant No. 2023YFA1606703, and by the National Natural Science Foundation of China under Grant Nos. 12435007 and 12361141819.


\begin{thebibliography}{99}

\bibitem{Denig:2012by}
A.~Denig and G.~Salme,
Prog. Part. Nucl. Phys. \textbf{68} (2013), 113-157.


\bibitem{Pacetti:2014jai}
S.~Pacetti, R.~Baldini Ferroli and E.~Tomasi-Gustafsson,
Phys. Rept. \textbf{550-551} (2015), 1-103.

\bibitem{Ping:2013jka}
R.~G.~Ping,
Chin. Phys. C \textbf{38} (2014), 083001.


\bibitem{BESIII:2017lkp}
M.~Ablikim \textit{et al.} [BESIII],
Chin. Phys. C \textbf{41} (2017), 063001.


\bibitem{BaBar:2007fsu}
B.~Aubert \textit{et al.} [BaBar],
Phys. Rev. D \textbf{76} (2007), 092006.

\bibitem{BESIII:2017hyw}
M.~Ablikim \textit{et al.} [BESIII],
Phys. Rev. D \textbf{97} (2018), 032013.


\bibitem{BESIII:2019nep}
M.~Ablikim \textit{et al.} [BESIII],
Phys. Rev. Lett. \textbf{123} (2019), 122003.

\bibitem{BESIII:2019cuv}
M.~Ablikim \textit{et al.} [BESIII],
Phys. Rev. Lett. \textbf{124} (2020), 032002.


\bibitem{BESIII:2020uqk}
M.~Ablikim \textit{et al.} [BESIII],
Phys. Lett. B \textbf{814} (2021), 136110.

\bibitem{BESIII:2021aer}
M.~Ablikim \textit{et al.} [BESIII],
Phys. Lett. B \textbf{820} (2021), 136557.

\bibitem{BESIII:2021rkn}
M.~Ablikim \textit{et al.} [BESIII],
Phys. Lett. B \textbf{831} (2022), 137187.

\bibitem{Belle:2022dvb}
G.~Gong \textit{et al.} [Belle],
Phys. Rev. D \textbf{107} (2023), 072008.

\bibitem{BESIII:2023ldb}
M.~Ablikim \textit{et al.} [BESIII],
Phys. Rev. D \textbf{109} (2024), 034029.

\bibitem{Iachello:1972nu}
F.~Iachello, A.~D.~Jackson and A.~Lande,
Phys. Lett. B \textbf{43} (1973), 191-196.

\bibitem{Iachello:2004aq}
F.~Iachello and Q.~Wan,
Phys. Rev. C \textbf{69} (2004), 055204.

\bibitem{Bijker:2004yu}
R.~Bijker and F.~Iachello,
Phys. Rev. C \textbf{69} (2004), 068201.

\bibitem{Yang:2019mzq}
Y.~Yang, D.~Y.~Chen and Z.~Lu,
Phys. Rev. D \textbf{100} (2019), 073007.

\bibitem{Li:2020lsb}
Z.~Y.~Li and J.~J.~Xie,
Commun. Theor. Phys. \textbf{73} (2021), 055201.

\bibitem{Dai:2021yqr}
A.~X.~Dai, Z.~Y.~Li, L.~Chang and J.~J.~Xie,
Chin. Phys. C \textbf{46} (2022), 073104.


\bibitem{Yan:2023nlb}
B.~Yan, C.~Chen, X.~Li and J.~J.~Xie,
Phys. Rev. D \textbf{109} (2024), 036033.


\bibitem{Chen:2023oqs}
C.~Chen, B.~Yan and J.~J.~Xie,
Chin. Phys. Lett. \textbf{41} (2024), 021302.

\bibitem{Chen:2024luh}
C.~Chen, B.~Yan and J.~J.~Xie,
Chin. Phys. C \textbf{49} (2025), 023102.

\bibitem{BaBar:2013ves}
J.~P.~Lees \textit{et al.} [BaBar],
Phys. Rev. D \textbf{87} (2013), 092005.

\bibitem{BESIII:2015axk}
M.~Ablikim \textit{et al.} [BESIII],
Phys. Rev. D \textbf{91} (2015), 112004.

\bibitem{BESIII:2019tgo}
M.~Ablikim \textit{et al.} [BESIII],
Phys. Rev. D \textbf{99} (2019), 092002.

\bibitem{BESIII:2019hdp}
M.~Ablikim \textit{et al.} [BESIII],
Phys. Rev. Lett. \textbf{124} (2020), 042001.

\bibitem{BESIII:2021rqk}
M.~Ablikim \textit{et al.} [BESIII],
Phys. Lett. B \textbf{817} (2021), 136328.

\bibitem{CMD-3:2015fvi}
R.~R.~Akhmetshin \textit{et al.} [CMD-3],
Phys. Lett. B \textbf{759} (2016), 634-640.

\bibitem{BESIII:2021tbq}
M.~Ablikim \textit{et al.} [BESIII],
Nature Phys. \textbf{17} (2021), 1200-1204.

\bibitem{Achasov:2014ncd}
M.~N.~Achasov \textit{et al.},
Phys. Rev. D \textbf{90} (2014), 112007.

\bibitem{SND:2022wdb}
M.~N.~Achasov \textit{et al.} [SND],
Eur. Phys. J. C \textbf{82} (2022), 761.

\bibitem{SND:2023fos}
M.~N.~Achasov \textit{et al.} [SND],
Phys. Atom. Nucl. \textbf{86} (2023), 1165-1172.

\bibitem{Li:2021lvs}
Z.~Y.~Li, A.~X.~Dai and J.~J.~Xie,
Chin. Phys. Lett. \textbf{39} (2022), 011201.

\bibitem{Yan:2023yff}
B.~Yan, C.~Chen and J.~J.~Xie,
Phys. Rev. D \textbf{107} (2023), 076008.

\bibitem{BESIII:2020nme}
M.~Ablikim \textit{et al.} [BESIII],
Chin. Phys. C \textbf{44} (2020), 040001.

\bibitem{BESIII:2017kqg}
M.~Ablikim \textit{et al.} [BESIII],
Phys. Rev. Lett. \textbf{120} (2018), 132001.

\bibitem{BESIII:2023rwv}
M.~Ablikim \textit{et al.} [BESIII],
Phys. Rev. Lett. \textbf{131} (2023), 191901.

\bibitem{Belle:2008xmh}
G.~Pakhlova \textit{et al.} [Belle],
Phys. Rev. Lett. \textbf{101} (2008), 172001.

\bibitem{Jia:2023upb}
S.~Jia, W.~Xiong and C.~Shen,
Chin. Phys. Lett. \textbf{40} (2023), 121301.



\end{thebibliography}
\end{document}